\documentstyle[12pt]{article}
\begin{document}

\baselineskip=18pt
\title{A Step Toward Pregeometry I.: \\
Ponzano--Regge Spin Networks \\
and the Origin of \\
Spacetime Structure \\
in Four Dimensions}
\author{Norman J. LaFave \\
Lockheed Engineering \& Sciences Company, \\
Houston, Texas 77058 \\
(Receipt}

\maketitle

In this paper, a candidate for pregeometry, Ponzano--Regge spin
networks, will be examined in the context of the pregeometric philosophy
of Wheeler. Ponzano and Regge were able to construct a theory for
$3$--dimensional quantum gravity based on $3nj$--symbols, obtaining the
path integral over the metric in the semiclassical limit. However,
extension of this model to $4$--dimensions has proven to be difficult.
It will be shown that the building blocks for $4$--dimensional spacetime
are already present in the Ponzano--Regge formalism using a
reinterpretation of the theory based on the pregeometric hypotheses of
Wheeler. \bigskip

PACS numbers: ***

\vfill \eject

\centerline{\bf I. THE NECESSITY OF PREGEOMETRY IN} \centerline{\bf
SPACETIME STRUCTURE AND QUANTUM GRAVITY} \medskip

        Wheeler${}^1$ points out a major deficiency of differential geometry
for the description of physics at the microscale---the modelling of the
dynamics of spin--$1/2$ particles in the microstructure. A structure
which claims to describe physics at the microscale must be judged to be
inadequate (or incomplete) if it does not have the ability to describe
spin--$1/2$ particles and their dynamics. The deficiency may be
described as follows.

        A $3$--manifold is not completely classified by the designation of its
topology, differential structure, and metric. Given a $3$--manifold with
the topology of a $3$--sphere endowed  with $n$ handles or wormholes,
$2^n$ different continuous fields of orthogonal triads ($2^n$ different
spin structures) may be admitted to the $3$--geometry which are
inequivalent to one another under any continuous sequence of small
readjustments${}^2$. These spin structures for a given $3$--geometry may
be distinguished from one another by a set of n descriptors $w_1$,
$w_2$,$\ldots$, $w_k$,$\ldots$, $w_n$ each of which take on the values
$+1$ or $-1$.

        The fact that we can now describe a spin--$1/2$ particle on a
$3$--geometry is not sufficient to describe the dynamics of these
particles. Processes such as particle pair creation require a change in
the topology of the $3$--geometry (the topology of the $3$--geometry
develops a new wormhole to accommodate the two new spin structures).
Classical differential geometry forbids such a topology change.
Therefore, it becomes clear that $3$--geometries, considered as
structures of classical differential geometry only, cannot be the basic
building blocks of the microstructure. Indeed, geometry as we know it
may need to give way to other structures at the extreme microscales.
This is not a new concept. Riemann${}^3$ suggested this possibility as
early as 1876.

        In addition to this deficiency, several practical difficulties in the
evaluation of the standard Hartle--Hawking path integral${}^4$ of
quantum gravity have not been dealt with in a satisfactory fashion
adding to the impression that the standard formalism may need to be
abandoned or altered significantly: \begin{itemize} \item The summation
over manifolds with different topologies has not been accomplished
because (1) a classification of all distinct topologies in four
dimensions has not been found${}^5$, and (2) the process for the
determination of the weight $\nu (M)$ for a given manifold is not known.

\item The transition from the Lorentzian path integral of quantum
mechanics to the Euclidean path integral is dubious. In flat space
quantum field theories, this ``Euclideanization" is performed via a
contour integral in the first quadrant of the complex time plane. The
lack of poles in the action inside the contour means that the two path
integrals may be equated. However, this ``Euclideanization" is only
assumed for the path integral of quantum gravity${}^6$. This is due to
the possible existence of singularities in the Einstein action on the
complex time plane within the contour for a general Euclidean metric.

\item The Euclidean action is not positive definite${}^7$. The conformal
invariance of the metric leads to infinite negative values for the
action for some metrics. This yields an infinite contribution to the path
integral. Attempts to eliminate this problem by the inclusion of $R^2$
terms${}^8$ in the action have been only partially successful and lead
to new questions concerning the physical nature of the new terms (i.e.
lack of unitarity). The splitting of the path integral into an integral
over conformal equivalence classes of $4$--geometries and an integral
over conformal factors and application of an assumed positive energy
conjecture yield some success${}^7$, but the positive energy conjecture
must be proven for general cases for this methodology to be believed.

\item The path integral is not perturbatively renormalizable${}^9$.

\item The interpretation of the concepts of time, observables, and
measurement in the covariant formalism appears to be ambiguous${}^{10}$.
\end{itemize} These problems have hindered the performance of any
reasonable calculations in quantum gravity despite a large number of
attempts to repair and/or understand them. They suggest a profound
deficiency in our understanding of the microstructure. \bigskip

\centerline{\bf A. Some Deep Questions} \medskip

These difficulties lead to several deep questions concerning quantum
gravity. Many of these questions are not answerable using the standard
path integral formalism: \begin{itemize} \item What is the role of time
in quantum gravity? Is time merely a semiclassical concept?

\item What is the role of the fundamental constants ($c$ and $\hbar$) in
the physics of the microscale? Why do the fundamental constants have the
relative values we observe?

\item Is there a structural breakdown of manifolds at the extreme
microscale? Do topological fluctuations and/or manifold fluctuations
and/or dimensionality fluctuations exist in quantum gravity? If so, how
are these types of fluctuations implemented in the theory? Is the
microscale described better in terms of a discrete structure than a
continuous manifold? If so, what is the nature of this discrete
structure?

\item Why is classical spacetime $4$--dimensional? Why does classical
spacetime have a locally Minkowski structure? What is the origin of
classical law?

\item What is the role of matter in quantum gravity? How can gravity be
unified with the other forces of nature?

\item If singularities are avoided by quantum gravity, how does this
avoidance manifest itself? \end{itemize} The difficulties presented
here, coupled with the failure to find satisfactory solutions to them
despite numerous attempts, lead to the unavoidable conclusion that the
development of a theory of quantum gravity must be approached from a
radically new direction. Pregeometry, as proposed by Wheeler${}^{11}$,
may be the key to defining that direction.

These questions are not treated as stumbling blocks in the construction
of pregeometry, but are utilized as a template for its construction.
Physical intuition based on past arguements concerning these questions,
along with some physically reasonable conjectures from general
observations, provide the basis for an approach to the development of
pregeometry.

In this paper we will examine a candidate for pregeometry,
Ponzano--Regge spin networks${}^{12}$, in light of the philosophy of
pregeometry. The concept of using spin networks to replace the spacetime
continuum at the microscale was first suggested by Penrose${}^{13}$ in
1972. Ponzano and Regge were able to construct a theory based on this
concept using the theory of $j$--symbols, but only for $3$--dimensional
quantum gravity. We will show that the building blocks for
$4$--dimensional spacetime are already present in the Ponzano--Regge
formalism using a reinterpretation of the theory based on the
pregeometric philosophy presented in the following section.

In Section II, the fundamental philosophy and hypotheses of Wheeler's
conception of pregeometry are reviewed. We shall discuss the
interpretation of these hypotheses and delineate those which play a
central role in the model discussed in this paper. In Section III, an
approach to the development of pregeometry based on the philosophy and
interpretations of Section II is outlined. The standard Ponzano--Regge
model which uses the interpretation of Hasslacher and Perry${}^{14}$ is
presented, both as an example of this approach in action, and as an
introduction to the basis on which the $4$--dimensional model is
constructed. In Section IV, the new pregeometric model in four
dimensions is presented with discussion of its properties and
implications. Finally, in Section V, a brief discussion of on--going
work and possible extensions of the work are outlined. \bigskip

\centerline{\bf II. THE WHEELERIAN CONCEPTION OF PREGEOMETRY} \medskip

Most attempts to solve the problems of quantum gravity have involved the
addition of new complexities to the model: the introduction of new terms
to the action ($R^2$ terms), the reinterpretation or extension of the
space of dynamical variables (spinors and twistor theory, Kaluza--Klein
Theory), or the introduction of a new type of structure (string theory).
In their early conceptions, many of these ``extended" theories have
shown great promise. However, at present, not one of these theories has
yet to produce answers to all of the questions we addressed in the last
section. In cases where a problem has been alleviated, a new problem or
set of problems has been generated. Furthermore, the additional
complexity of these theories have produced a new set of intractable
problems in the solution of the equations. The perturbation methods,
which have been so successful in the solution of other quantum field
theories, have failed due to the lack of a finite perturbation expansion
at the Planck scale${}^9$. This is due to the dimensional character of
the gravitational constant.

The pregeometry of Wheeler takes a fundamentally different approach. It
assumes that fundamental laws of nature should be based on simple
principles. Indeed, it is the belief of Wheeler that the fundamental
laws of nature should be synthesized from no law at all. Pregeometry, in
its present conception, is not a formal, rigorous theory, but a set of
hypotheses regarding the fundamental nature of the physical world. They
are based on general observations of the basic properties of quantum
systems. A summary of some of the basic hypotheses of Wheelerian
pregeometry is contained in Wheeler's ``three questions", ``four no's",
and ``five clues". \bigskip

\centerline{\bf A. The Three Questions} \medskip

The ``Three Questions" posed by Wheeler may seem to be philosophical in
nature. However, they are at the heart of Wheeler's attempts to gain
insight into the foundations of quantum phenomenon. \begin{enumerate}
\item {\it How come existence?}

What does it mean to say that an object exists or that a process has
taken place? Existence seems built from the choosing of yes--no
questions by the observer--participants in the universe and the
eliciting of answers through irreversible measurements by them. The
click of a counter, the deflection of a needle in a meter, or the
registration of an interference pattern on a screen all exhibit an
information--theoretic (yes/no) nature. In the case of the counter the
information is dealt with in one yes/no bit of information (Does the
counter count?). In the case of the needle in the meter or the
interference pattern, large numbers of bits of information are gathered
(charged particles moving through a wire or particles registering on the
screen). Wheeler refers to this as ``It from Bit". The universe,
according to Wheeler, is participatory in nature and defined via
discrete bits of information. What we refer to as ``facts" (a counter
clicks, a needle deflects) are dependent on the questions we choose to
ask and the answer we receive.

\item {\it How come the quantum?}

Why does the world (existence) exhibit this information theoretic nature
via the quantum measurement process? Meaning seems built on a series of
information-theoretic yes-no questions and resulting observations by the
many observer--participants. Why?

\item {\it How come ``one world" out of many observer--participants?}

How does the one world we live in arise out of the gathering of
information by the many observer--participants? \end{enumerate}

To provide ourselves a foundation for answering these questions, Wheeler
adopted ``four no's" for guidance. \bigskip

\centerline{\bf B. The Four No's} \medskip

\begin{enumerate} \item {\it No Tower of Turtles}

Wheeler states that there is no structure built on a structure built on
yet another structure, ad infinitum. Instead, a feedback loop, the
meaning circuit, is adopted as the fundamental nature of the
observer--participant world. In the meaning circuit, physics gives rise
to observer--participancy, which gives rise to information which defines
physics (see Fig. 1).

\item {\it No Laws}

The laws of physics exist, but from whence do they come? Why do they
have their unique functional form? Nothing in our present formalisms
have given direct clues as to the origin of the classical field
equations. It is clear that the laws of physics cannot have existed from
the beginning of time. They must have come into existence at the big
bang. It seems that the only answer is to assume that they are produced
from no law at all.

This is probably the hardest conception in Wheeler's set of hypotheses.
Conceiving of something as elegant as physical law evolving from nothing
is difficult. However, one may envision this difficult hypothesis by
imagining a ``space" of all consistent mathematical structures and laws.
This all--encompassing set of possible physical laws and structures is
as good as no law and no structure at all. The observer--participants
``pick" out those structures and laws which define physics by their
observer--participancy via the meaning circuit (see Fig. 2).

\item {\it No Continuum}

Mathematical logic has provided no evidence of the existence of a number
continuum, despite decades of research. The use of irrational numbers
simplifies the laws of arithmetic in a convenient fashion, but their
existential nature is purely a mythology. Wheeler states, ``Nothing so
much distinguishes physics as conceived today from mathematics as the
difference between the continuum character of the one and the discrete
character of the other". However, belief in the bit--wise,
information--theoretic character of quantum physics closes this gap.
Therefore, the continuous manifold of classical physics becomes merely a
convenient artifice which simplifies the form of the classical physical
laws, but has no existential character.

Indeed, recent work in the loop representation of general
relativity${}^{10,15}$ has produced evidence that the structures which
give us manifolds at the classical level, must be discrete at the
microscale (discrete areas for the loops which yield classical
structures).

\item {\it No Space and No Time}

Space and time are not physical entities but conceptions, created by
man, to keep track of the order of things---spatial and temporal.
Quantum theory imposes fluctuations on spacetime that, at the Planck
scale, deprive spacetime of its connectivity and space and time of their
classical meaning. Time and space must be derived in the classical limit
of our information--theoretic universe. \end{enumerate}

We will now examine ``five clues" based on general observations about the
operation of quantum mechanics. \bigskip

\centerline{\bf C. The Five Clues} \medskip

\begin{enumerate} \item {\it Austerity---The Boundary of a Boundary is
Zero}

This principle of algebraic topology, the boundary of a boundary is zero
(BB principle), is a simple, tautological principle. However, it is also
a unification principle of almost every modern field theory as will be
exhibited later in this paper. From this simple principle, is derived
all of the conservation laws in field theory and relativity. As we shall
see in Section IV, it is not the only austerity principle with this kind
of general applicability.

\item {\it Observer--Participancy---No Question, No Answer}

The choice of question and when it is asked plays a part in determining
what is physical fact. The location or momentum of an electron are
completely unknown until they are measured. Furthermore, upon measuring
one of these quantities to high accuracy, the other cannot be measured
by any measurement process. This is the very essence of Bohr's
Copenhagen interpretation of quantum mechanics.

\item {\it Timelessness---The Super--Copernican Principle}

There is no ``now--centeredness". Meaning is the union of all
information generated by observer--participants which communicate,
regardless of time. The observer--participants to come play a role in
the establishment of the ``bit--based reality" of today by the questions
they will choose to ask and the answers they will receive, just as the
observer--participants of the past and present contribute to defining
this reality.

\item {\it One world from many Observer--Participants---``Consciousness"}

Since consciousness is an ill--defined concept, we avoid it in our
establishment of meaning for the present. We concentrate on
communication and the establishment of meaning regardless of the details
of the communicators.

\item {\it Self--synthesis---More is Different}

Large numbers of entities generate new phenomena which do not exist for
samll numbers of entities. For example, large numbers of water molecules
in a container exhibit gas liquid and solid phases. Furthermore, large
numbers of Helium--3 atoms exhibit superfluidity. Superconductivity is
another example.

\end{enumerate}

The full conception of pregeometry as envisioned by Wheeler, information
theoretical world defined by observer--participancy, is not derived here.
This is a difficult task and is beyond our present understanding of
nature. We shall, in lieu of this daunting task, use three of the basic
philosophies of pregeometry as a template for deriving a consistent
theory of quantum gravity.

These philosophies may be stated succinctly as: 1.) Physical principles
and structures should become more austere and general as they become
more fundamental; 2.) Time and space, as we conceive of them in our
classical world, have no place in the description of the microscale
structure; 3.) The continuum is a classical artifice and has no
existential character. The ``pregeometry" derived in this paper uses
these philosophies as its foundation. \bigskip

\centerline{\bf III. AN APPROACH TO THE CONSTRUCTION OF} \centerline{\bf
PREGEOMETRIC QUANTUM GRAVITY} \medskip

\centerline{\bf A. A Description of the} \centerline{\bf Construction of
Pregeometry } \medskip

        For the purpose of this paper, we will redefine pregeometry as a set of
primitive building blocks and simple rules which lead to geometric
structure in an appropriate limit. Pregeometry, defined in this manner,
certainly cannot be deduced from the study of classical geometrodynamics
anymore than atomic and molecular structure can be explained by the
study of elasticity. It is a system which is less constrained (more
austere and more fundamental) than geometrodynamics. However, a
knowledge of the physics of classical geometrodynamics may be used as a
guide to the development of a consistent pregeometry just as a knowledge
of elasticity may be used as a guide to the development of consistent
theories of atomic and molecular structure. It is this methodology which
we employ in the development of an approach to pregeometric structure.

        The approach we use to construct pregeometry may be described in
general by the following algorithm: \begin{enumerate} \item Use
``physical intuition'' to construct a quantum theory (pregeometry) from
building blocks based on a set of fundamental constants with some simple
rules of interaction between the building blocks. The pregeometry should
not, by definition, have a purely geometrical structure.

        This step requires conjecture. However the ``physical intuition''
(based on the general observations presented in section I and the
hypotheses of section II) used in the construction will help to improve
our chance of success.

\item Define a sensible semiclassical limit in terms of the set of
fundamental constants using physical intuition. The resulting structure
is compared with our knowledge of the physics of spacetime (Is the
structure a  reasonable representation of spacetime in the classical
limit? Are the semiclassical equations for the structure recognizable as
relativistic quantum mechanical expressions?). If a reasonable
semiclassical structure is not obtained (does not have a recognizable
form), or if the semiclassical theory is not consistent, restart the
development at step 1 with a quantum theory built from different
building blocks with different rules of interaction.

\item Determine the physical interpretation of the role of the
fundamental constants from the comparison implemented in step 2.
\end{enumerate}

        We shall call this algorithm the {\it pregeometry construct
methodology} (PCM). The basic purpose of the PCM is to obtain geometry
and its dynamics as the limit of a structure which is primitive and
essentially nongeometrical. The hope is that there is only a small
number of quantum structures which may be formulated using the
methodology of step 1 (preferably one) which have a consistent and
physical semiclassical limit. The PCM is summarized by Fig. 3. \bigskip

\centerline{\bf B. Example of the Approach:} \centerline{\bf The
Standard Ponzano--Regge Quantum Gravity} \medskip

        To exhibit the functionality of the PCM for the construction of
pregeometry, we exhibit a theory, constructed using the PCM, which yields
$3$--dimensional quantum gravity in the semiclassical limit: the
standard Ponzano--Regge model${}^{12}$. This model is a ``toy model'' of
quantum gravity. However, it provides an excellent example of the use of
the PCM  and will be the basis for the model presented in Section IV of
this paper. The meaning of the abstract expression ``physical
intuition'' used in the description of the PCM in the last subsection
will become apparent from the following discussion. \bigskip

\centerline{\bf 1. Description of Ponzano--Regge Quantum Gravity}
\medskip

        Consider a $3nj$--symbol description ($n$ large) of a closed
interaction of half--integral spins whose diagramatic representation is
given by a closed, finely triangulated $2$--surface with $2n$ triangles
and $3n$ edges (we use the diagramatic method of Ponzano and Regge which
is the surface dual to the diagramatic method of Yutsis et.
al.${}^{16}$). The fineness of the diagram, determined by $n$, helps
define the precision of the geometry derived from the resulting
pregeometry. This triangulated $2$--surface has a total spin value of
$(m+\frac{1}{2})\hbar$ assigned to each of the edges of the diagram,
where $m$ has a value from the sequence
$0,\frac{1}{2},1,\frac{3}{2},2,\ldots\,\,$. The rules for the addition
of spin are satisfied by the three edges of each triangle of the
$2$--surface (each vertex of a diagram from the method of Yutsis et.
al.). In other words, the spin interaction at the triangle represented
by Fig. 4 is given by a Wigner $3j$--symbol represented by
\begin{equation} \left( \begin{array}{ccc} a & b & c \\ \alpha & \beta &
\gamma \end{array} \right) \, , \label{eq-3jsymbol} \end{equation} which
is evaluated using the expression \begin{eqnarray} \lefteqn{ \left(
\begin{array}{ccc} a & b & c \\ \alpha & \beta & \gamma \end{array}
\right) = (-1)^{a - b - \gamma}} \hspace{1.0in} \nonumber \\
    & & \nonumber \\
    & & \times [ \triangle (abc) \, (a + \alpha)! \, (a - \alpha)! \, (b
+ \beta)! \, (b - \beta)! \, (c + \gamma)! \, (c - \gamma)!
]^{\frac{1}{2}} \nonumber \\
    & & \nonumber \\
    & & \times \sum_x (-1)^x \, [ x! \, (c - b + \alpha + x)! \, (c - a
- \beta + x)! \nonumber \\
    & & \nonumber \\
    & & \times (a + b - c - x)! \, (a - \alpha - x)! \, (b + \beta - x)!
]^{-1} \, , \label{eq-3jsymbexp} \end{eqnarray} where $a$,$b$, and $c$,
the total spin quantum numbers, are integer or half--integer, and
$\alpha$, $\beta$, and $\gamma$ are the quantum numbers for the
components of the spin along a given fixed direction, corresponding to
the spin quantum number $a$, $b$, and $c$, respectively. The factor
$\triangle(abc)$ is given by the expression \begin{equation}
\triangle(abc) = \frac{(a+b-c)! \, (b+c-a)! \, (c+a-b)!}{(a+b+c+1)!} \, .
\label{eq-triabc} \end{equation} This factor is zero unless the
variables $a$,$b$, and $c$ satisfy the triangle inequalities
\begin{equation} \vert a-b \vert \leq c \leq a+b \, , \label{eq-triineq1}
\end{equation} \begin{equation} \vert a-c \vert \leq b \leq a+c \, ,
\label{eq-triineq2} \end{equation} and \begin{equation} \vert b-c \vert
\leq a \leq b+c \, . \label{eq-triineq3} \end{equation} Furthermore, the
relation $\alpha+\beta+\gamma=0$ must be satisfied, and
$a-\vert\alpha\vert$, $b-\vert\beta\vert$, and $c-\vert\gamma\vert$ must
be natural integers. If any of these conditions are not satisfies, the
$3j$--symbol is considered to be equal to zero.

        The simplest example of a  $3nj$--symbol is the $6j$--symbol, which is
given as a sum of products of four Wigner $3j$--symbols by the expression
\begin{eqnarray} \lefteqn{ \left\{ \begin{array}{ccc} a & b & c \\ d & e
& f \end{array} \right\} =  \sum_{{\alpha \, \beta \, \gamma} \atop
{\delta \, \epsilon \, \phi}} (-1)^{d+e+f+\delta+\epsilon+\phi}}
\hspace{0.8in} \nonumber \\
    & & \nonumber \\
    & & \times \left( \begin{array}{ccc} d & e & c \\ \delta & -\epsilon
& \gamma \end{array} \right) \left( \begin{array}{ccc} e & f & a \\
\epsilon & -\phi & \alpha \end{array} \right) \left( \begin{array}{ccc}
f & d & b \\ \phi & -\delta & \beta \end{array} \right) \left(
\begin{array}{ccc} a & b & c \\ \alpha & \beta & \gamma \end{array}
\right) \, . \label{eq-sixjfromwigner} \end{eqnarray} Note that the
quantum numbers for the components of the spins along a fixed direction
have been eliminated by a summation over there allowed values.
Expression (\ref{eq-sixjfromwigner}) has been evaluated by
Racah${}^{12}$ to obtain the combinatorial expression \begin{eqnarray}
\lefteqn{ \left\{ \begin{array}{ccc} a & b & c \\ d & e & f \end{array}
\right\} =  [\triangle(abc) \, \triangle(aef) \, \triangle(cde) \,
\triangle(bdf)]^{\frac{1}{2}} \, \sum_x (-1)^x \, (x+1)!} \hspace{0.8in}
\nonumber \\
    & & \nonumber \\
    & & \times [(a+b+d+e-x)! \, (a+c+d+f-x)! \, (b+c+e+f-x)! \nonumber \\
    & & \nonumber \\
    & & \times (x-a-b-c)! \, (x-a-e-f)! \nonumber \\
    & & \nonumber \\
    & & \times (x-c-d-e)! \, (x-b-d-f)!]^{-1} \, . \label{eq-sixjracah}
\end{eqnarray} This $6j$--symbol has a diagramatic representation given
by the tetrahedron in Fig. 5.  The $6j$--symbol vanishes if any one of
the triangles $(abc)$, $(cde)$, $(aef)$, or $(bdf)$ violate the triangle
inequalities, or if any of the sums $q_1 =a+b+c$, $q_2=a+e+f$,
$q_3=b+d+f$, $q_4=c+d+e$, $p_1=a+b+d+e$, $p_2=a+c+d+f$, or $p_3=b+c+e+f$
are not integers ($p_h \geq q_k$; $h,k=1,2,3,4$ by the triangle
inequalities).

        Now, let us return to the consideration of the diagramatic
representation of a general $3nj$--symbol as a triangulated
$2$--surface. We shall denote the $3nj$--symbol by $[D]$ and the
corresponding $2$--surface diagram by $D$. The evaluation of $[D]$ may
be performed by expanding it as a sum of products of $p$ $6j$--symbols
which are denoted by $[T_k]$. This expansion corresponds to the filling
of the interior of the closed, triangulated $2$--surface $D$ with $p$
tetrahedra such that there are $q$ internal edges $x_1,\ldots,x_q$ which
are summed over their allowed values. This filling procedure yields a
new set of $3$--dimensional diagrams ${\cal D}^{p,q}_l(D)$ with every
diagram element of the set, called a spin network configuration
(distinguished by an integer $l$), having a fixed number of tetrahedra
$p$, fixed number of internal edges $q$, fixed connectivity, and the
internal edges $x_1,\ldots,x_q$ have fixed lengths. All possible
combinations of the allowed values of the internal edges
$x_1,\ldots,x_q$ determine all of the spin network configurations of the
set. It should be noted that the filling procedure (connectivity and the
number $p$ of tetrahedra) used to construct ${\cal D}^{p,q}_l(D)$ does
not effect the value of $[D]$. Only the $2$--surface triangulation and
the $3$--topology affect the value of the $3nj$--symbol. The expansion
to determine $[D]$ is given by \begin{equation} [D] = \sum_{x_1} \cdots
\sum_{x_q} A(x_1,\ldots,x_q) \, , \label{eq-3nj} \end{equation} where,
for fixed values of the internal edge lengths $x_1,\ldots,x_q$,
\begin{equation} A(x_1,\ldots,x_q) = \prod^p_{k=1} \, [T_k] \,
(-1)^{\chi} \, \prod^q_{i=1} \, (2x_i+1) \, , \label{eq-3njweight}
\end{equation} and where, if $D$ and ${\cal D}^{p,q}_l(D)$ are
homeomorphic to a $2$--sphere and a $3$--ball, respectively, then
\begin{equation} \chi = \sum^q_{j=1} \, (n_j-2) \, x_j + \chi_0 \, .
\label{eq-chi} \end{equation} The variable $n_j$ denotes the number of
tetrahedra belonging to $x_j$, and $\chi_0$ is a fixed phase chosen in
order to make $\chi$ an integer. $A(x_1,\ldots,x_q)$ is the weight of a
spin network configuration of ${\cal D}^{p,q}_l(D)$.

        If no internal vertices are placed within $D$ to obtain the set ${\cal
D}^{p,q}_l(D)$, the expression (\ref{eq-3nj}) is finite and $[D]$ may be
determined directly. If internal vertices are placed within $D$ to
obtain the set ${\cal D}^{p,q}_l(D)$, the expression (\ref{eq-3nj}) is
infinite and requires a renormalization procedure. The renormalization
procedure is described as follows. First, consider the simple case
described by Fig. 6. $[D]$ is a $6j$--symbol, and it is subdivided into
four $6j$--symbols by placing a single vertex within the interior of the
diagram $D$ and connecting four interior edges from this interior vertex
to the four vertices of $D$. For this case, expression (\ref{eq-3nj})
yields the expression \begin{eqnarray} \lefteqn{ [D] = \left\{
\begin{array}{ccc} a & b & c \\ d & e & f \end{array} \right\} =}
\hspace{.25in} \nonumber \\
    & & \nonumber \\
    & &  \sum_{allowed\,domain\,of\,x,y,z,t} (2x + 1)(2y + 1)(2z + 1)(2t
+ 1) \, (-1)^{x+y+z+t+a+b+c+d+e+f} \hspace{.25in} \nonumber \\
    & & \nonumber \\
    & & \times \left\{ \begin{array}{ccc} a & b & c \\ x & y & z
\end{array} \right\} \left\{ \begin{array}{ccc} f & e & a \\ y & z & t
\end{array} \right\} \left\{ \begin{array}{ccc} d & b & f \\ z & t & x
\end{array} \right\} \left\{ \begin{array}{ccc} c & e & d \\ t & x & y
\end{array} \right\} \, . \label{eq-sixjidentity} \end{eqnarray} Using
the Elliot--Biedenharn Identity${}^{17}$ derived from the diagram in
Fig. 7 \begin{eqnarray} \lefteqn{  \left\{ \begin{array}{ccc} g & h & j
\\ e & a & d \end{array} \right\} \left\{ \begin{array}{ccc} g & h & j
\\ e' & a' & d'         \end{array} \right\} = \sum_x (-1)^{\phi} \, (2x + 1) }
\hspace{0.5in}  \nonumber \\
    & & \nonumber \\
    & & \times \left\{ \begin{array}{ccc} a & a' & x \\ d' & d & g
\end{array} \right\} \left\{ \begin{array}{ccc} d & d' & x \\ e' & e & h
\end{array} \right\} \left\{ \begin{array}{ccc} e & e' & x \\ a' & a & j
\end{array} \right\} \, , \label{eq-beidentity} \end{eqnarray} where
\begin{equation} \phi = g+h+j+e+a+d+e'+a'+d'+x \, , \label{phiexp}
\end{equation} and $x$ runs over an allowed domain, on the summation
over $t$ in expression (\ref{eq-sixjidentity}), we obtain
\begin{equation} [D] =  \left\{ \begin{array}{ccc} a & b & c \\ d & e &
f \end{array} \right\} \, \sum_{xyz} (2x+1)(2y+1)(2z+1) \left\{
\begin{array}{ccc} a & b & c \\ x & y & z \end{array} \right\}^2 \, .
\label{eq-sixj2} \end{equation} Applying the identity \begin{equation}
\sum_x (2x+1) \, \left\{ \begin{array}{ccc} a & b & x \\ d & e & f
\end{array} \right\} \left\{ \begin{array}{ccc} a & b & x \\ d & e & f'
\end{array} \right\} =\frac{\delta_{ff'}}{2f+1} \, ,
\label{eq-deltaident} \end{equation} to the sum over $z$ in expression
(\ref{eq-sixj2}) yields \begin{equation} [D] =  \left\{
\begin{array}{ccc} a & b & c \\ d & e & f \end{array} \right\} \,
\sum_{xy} \frac{\delta_{xyc}}{2c+1} \, (2x+1)(2y+1) \, , \label{eq-sixj3}
\end{equation} and since \begin{equation} \sum^{y=x+c}_{y= \vert x-c
\vert} \, (2y+1) = (2x+1)(2z+1) \, , \label{eq-simpidentity}
\end{equation} expression (\ref{eq-sixj3}) becomes \begin{equation} [D]
=  \left\{ \begin{array}{ccc} a & b & c \\ d & e & f \end{array}
\right\} \, \sum^{\infty}_{x=0} \, (2x+1)^2 \, , \label{eq-sixj4}
\end{equation} which is infinite due to the summation. However, let us
limit the summation on $x$ up to $x=R$, where $R$ is large. We find that
\begin{equation} {\cal R}(R) \equiv \sum^R_{x=0} (2x+1)^2 \simeq
\frac{4R^3}{3} \, . \label{eq-renormfac} \end{equation} This result
implies that the renormalized expression for the $6j$--symbol $[D]$ is
given by \begin{eqnarray} \lefteqn{ [D] = \left\{ \begin{array}{ccc} a &
b & c \\ d & e & f \end{array} \right\} = \lim_{R \rightarrow \infty} \,
({\cal R}(R))^{-1}} \nonumber \\
    & & \nonumber \\
    & & \times \sum_{allowed\,domain\,of\,x,y,z,t \, < \, R} (2x + 1)(2y
+ 1)(2z + 1) (2t + 1) \, (-1)^{x+y+z+t+a+b+c+d+e+f} \nonumber \\
    & & \nonumber \\
    & & \times \left\{ \begin{array}{ccc} a & b & c \\ x & y & z
\end{array} \right\} \left\{ \begin{array}{ccc} f & e & a \\ y & z & t
\end{array} \right\} \left\{ \begin{array}{ccc} d & b & f \\ z & t & x
\end{array} \right\} \left\{ \begin{array}{ccc} c & e & d \\ t & x & y
\end{array} \right\} \, . \label{eq-sixjidentren} \end{eqnarray} We note
that, for fixed x, the summation vanishes if $z>x+b$, $y>x+c$, or
$t>x+d$. Therefore, if $x<\min (R-b,R-c,R-d)$, then the limitations
$y,z,t<R$ have no effect on the summation and we may assume expression
(\ref{eq-sixjidentren}) to be correct. The renormalized expression for a
general $3nj$--symbol $[D]$, obtained using a filled diagram set ${\cal
D}^{p,q}_l(D)$ with $P$ internal vertices, is given, using a similar
procedure, by \begin{equation} [D] = \lim_{R \rightarrow \infty} \,
({\cal R}(R))^{-P} \, \sum_{x_1} \cdots \sum_{x_q} A(x_1,\ldots,x_q) \, ,
\label{eq-3njren} \end{equation} where $A$ is given by expression
(\ref{eq-3njweight}).

        The quantum structure represented by the filled diagram set ${\cal
D}^{p,q}_l(D)$ of our $3nj$--symbol $[D]$ has the characteristics
expected from step 1 of the PCM. \begin{enumerate} \item The building
blocks of the structure, half--integral spins, are built from the
fundamental constant $\hbar$.

\item The structure is basically non--geometrical. The $3$--geometries
are a subset of the set of all spin network configurations. Although the
triangle inequalities of any spin network configuration of ${\cal
D}^{p,q}_l(D)$ must be satisfied, the tetrahedral inequalities need not
be satisfied by the tetrahedral representation of nonvanishing
$6j$--symbols within  a spin network configuration of ${\cal
D}^{p,q}_l(D)$ (see Appendix A). Treating the spin value
$(m+\frac{1}{2})\hbar$ as a ``length'', Cayley's determinant,
\begin{equation} 2^3 \, (3!)^2 \, V^2 = \left\vert \begin{array}{ccccc}
0 & j^2_{34} & j^2_{24} & j^2_{23} & 1 \\ j^2_{34} & 0 & j^2_{14} &
j^2_{13} & 1 \\ j^2_{24} & j^2_{14} & 0 & j^2_{12} & 1 \\ j^2_{23} &
j^2_{13} & j^2_{12} & 0 & 1 \\ 1 & 1 & 1 & 1 & 0 \end{array} \right\vert
\, , \label{eq-cayley} \end{equation} may be used to determine the
``volume'' $V$ of the tetrahedral diagram of a given $6j$--symbol given
by Fig. 8, where $j_{kl}$ are the ``lengths'' of the edges of the
tetrahedral diagram. Two cases present themselves from the set of
possible nonvanishing $6j$--symbols: \begin{itemize} \item Case (A): The
$6j$--symbols have tetrahedral diagrams satisfying the tetrahedral
inequalities ($V^2>0)$. This set of $6j$--symbols (diagrams are
Euclidean tetrahedra) lead to our differential geometric structure.

\item Case (B): The $6j$--symbols have tetrahedral diagrams which
violate the tetrahedral inequalities ($V^2<0)$ (see Appendix A).
According to Hasslacher and Perry${}^{14}$, these tetrahedra (hyperflat
tetrahedra) correspond to a discrete foam--like structure. \end{itemize}
\end{enumerate}

        The physical intuition implemented in the construction of this spin
network pregeometry manifests itself in the guise of the ``geometric''
diagramatic representation of spin networks. Furthermore, the importance
of geometric relations in the pregeometry, such as tetrahedral volume,
and the triangle inequalities, yield additional insight into how
``geometry'' forms from this more primitive structure.

        The implementation of step 2 (the semiclassical limit) of the PCM
involves letting $\hbar$ go to zero (which is the same as letting
$j_{kl}$ of the $3nj$--symbol become large) and letting the number of
internal tetrahedra $p$ in the set ${\cal D}^{p,q}_l(D)$ for the
$3nj$--symbol become large. First, consider the expression for a case
(A) $6j$--symbol (see Fig. 5) when $j_{kl}$ becomes large
\begin{equation} \left\{ \begin{array}{ccc}
    a & b & c \\
    d & e & f \end{array} \right\} \approx { \left( \frac{\hbar^3}{12
\pi V} \right) }^{\frac{1}{2}} \cos \left( \left( \sum^4_{k,l=1}
\frac{1}{\hbar} \, j_{kl} \, \theta_{kl} \right) + \frac{\pi}{4} \right)
\, , \label{eq-sixjasymptota} \end{equation} where $V$ is the volume of
the tetrahedron evaluated using Cayley's determinant (expression
(\ref{eq-cayley})), and $\theta_{kl}$ is the angle between the outward
unit normals to the faces that are separated by edge $kl$. The angle
$\theta_{kl}$ is evaluated using the expression \begin{equation} \sin
\theta_{kl} = \frac{3}{2} \, \frac{V \, j_{kl}}{A_k \, A_l} \, ,
\label{eq-sinthetaeuc} \end{equation} where $A_k$ and $A_l$ are the
areas of the triangles opposite to the vertices $k$ and $l$,
respectively. Secondly, consider the expression for a Case (B)
$6j$--symbol (see Fig. 5) when $j_{kl}$ becomes large \begin{equation}
\left\{ \begin{array}{ccc}
    a & b & c \\
    d & e & f \end{array} \right\} \approx \pm { \left(
\frac{\hbar^3}{48 \pi \vert V \vert} \right) }^{\frac{1}{2}} \exp \left(
- \frac{1}{\hbar} \left\vert \sum^4_{k,l=1} \, j_{kl} \, Im \,
\theta_{kl} \right\vert \right) \, , \label{eq-sixjasymptotb}
\end{equation} where $\vert V \vert$ is the norm of the volume of the
tetrahedron evaluated using Cayley's determinant (expression
(\ref{eq-cayley})), and the expression for $\theta_{kl}$ is defined by
analytic continuation of the expression for $\theta_{kl}$ given in Case
(A) (expression (\ref{eq-sinthetaeuc})) \begin{equation} \theta_{kl} = m
\pi + i \, Im \, \theta_{kl}\, \,; \, \, \, \, m \, \, integer \, ,
\label{eq-comhypang} \end{equation} and \begin{equation} \sinh (Im \,
\theta_{kl}) = \cos (Re \, \theta_{kl}) \, \frac{3 j_{kl} \, \vert V
\vert}{2 A_k \, A_l} \, . \label{eq-sinthetaflat} \end{equation} $A_k$
and $A_l$ are the areas of the triangles opposite to the vertices $k$
and $l$, respectively, as before. We note that if $j_{kl}$ becomes
large, the damped exponential in expression (\ref{eq-sixjasymptotb})
causes the Case (B) $6j$--symbols to have no weight relative to most of
the Case (A) $6j$--symbols. Therefore, substituting expression
(\ref{eq-sixjasymptota}) into expression (\ref{eq-3nj}) through
expression (\ref{eq-3njweight}), noting that the summations in
expression (\ref{eq-3nj}) become integrals in the limits $\hbar
\rightarrow 0$ and $p \rightarrow \infty$, and assuming that the number
of tetrahedra $p$ is large enough to be a multiple of four and that the
action is small (near flat space), we obtain, after some manipulation
\begin{eqnarray} \lefteqn{[D] = \int_{D} \, dx_1 \cdots dx_q \, \Biggl(
\left( \frac{\hbar^3}{12 \pi} \right)^{\frac{p}{2}} \, (V_1 \cdots
V_p)^{-\frac{1}{2}} \, (1-i \, \exp(-2 \pi \, i \, N / p))^p}
\hspace{1.0in} \nonumber \\
    & & \nonumber \\
    & & \times \prod^q_{j=1} \, (2x_j+1) \Biggr) \, \, \exp \, \left[ -
S_R \, \left( \frac{\cos 2 \pi N / p}{1 + \sin 2 \pi N / p} \right)
\right] \, , \label{eq-3dpathint} \end{eqnarray} where $S_R$ is the
action for $3$--dimensional Regge calculus${}^{18}$ \begin{equation} S_R
= \sum^q_{j=1} \, x_j \left[ 2 \pi - \sum^{n_j}_{i=1} \, ( \theta^i_{j} -
\pi) \right] \, , \label{eq-reggeact} \end{equation} ($\theta^i_{j}$ is
the angle between the two outward unit normals to the triangles of
tetrahedron $i$ which meet at edge $j$, and $n_j$ is the number of
tetrahedra belonging to $j$) and $N = \sum^q_{j=1} \, x_j \, n_j$. The
factor between brackets in expression (\ref{eq-reggeact}) is the deficit
angle at the edge $j$. Expression (\ref{eq-3dpathint}) resembles the path
integral for $3$--dimensional simplicial quantum gravity. If $N/p$ is an
integer, then expression (\ref{eq-3dpathint}) is exactly the path
integral. If $N/p$ is not an integer, then the factor in parentheses
within the exponential is absorbed into the Newton's gravitational
constant, which makes the constant a function of the filling of $D$ to
form the set ${\cal D}^{p,q}_l(D)$.

        We note that step 3 of the PCM is easily accomplished. Comparison of
the form of the path integral, given by expression (\ref{eq-3dpathint}),
with the standard expression for the path integral associates the Regge
calculus edge length with the spin value $(m+\frac{1}{2})\hbar$ of an
edge, in agreement with the physical intuition stated earlier. In units
of $G=c=1$, the Planck length is just $\hbar = 2.63 \times 10^{-66}$.
The diagrammatic representation of the Ponzano--Regge model interpreted
in terms of the PCM is given by Fig. 9. \bigskip

\centerline{\bf 2. Properties of Ponzano--Regge Quantum Gravity} \medskip

        The pregeometry of Ponzano and Regge has several attractive properties,
many of which we would expect our physical pregeometry to exhibit. These
properties may be summarized as follows: \begin{enumerate} \item The
pregeometry contains a built--in renormalization cut--off: $\frac{1}{2}
\, \hbar$ (one--half of the Planck length).

\item The fact that the edge lengths of the diagrams vary in integer
units of $\frac{1}{2} \, \hbar$, and no length is ever zero, means that
no tetrahedral diagram will ever have a volume equal to zero (see
Appendix).

\item The discrete theory is a consistent generalization of the continuum
theory.

\item The fractalization procedure designated by the expression
(\ref{eq-sixjidentren}) allows the analyst to obtain an arbitrarily fine
tesselation of the interior of $D$ without changing the value of the
$3nj$--symbol $[D]$. Therefore, the lattice model at a fixed point is the
renormalization group transformation. The theory contains a natural
renormalization parameter given by the number of tetrahedra $p$ in the
diagram set ${\cal D}^{p,q}_l(D)$. This choice of renormalization
parameter differs from the one defined by Kadanoff${}^{19}$ where the
length scale is continuously rescaled.

\item Foam--like structures are automatically included in the theory by
the relaxation of the tetrahedral inequalities according to the
interpretation of Hasslacher and Perry${}^{14}$.

\item Calculations can be done without the evaluation of path integrals
since the general expression for the $3nj$--symbol is strictly
combinatorial in nature.

\item The theory of $3$--dimensional quantum gravity is reduced to simple
building blocks with simple laws: spins and their addition.

\item The tetrahedral inequalities and the weight $A$ tend to suppress
any spin network configuration of the set ${\cal D}^{p,q}_l(D)$, with a
finite number of tetrahedra $p$, which contains an edge or set of edges
that have infinite length relative to their neighbors. This allows the
avoidance of curvature singularities in a given diagram configuration
with finite $p$. We shall return to this point later. \end{enumerate}

        As was stated earlier, the Ponzano--Regge pregeometry has been treated
as nothing more than a ``toy model'' in its early development. However,
its unique and remarkable properties provided overwelming impetus to
adapt it to the construction of the physical pregeometry which yields
$4$--dimensional spacetime in the classical limit. Early attempts to
discover a direct extension to four dimensions were unsuccessful and the
model received only cursory study for some time${}^{11,14}$. There is
now renewed interest in this model due to four important developments:
1.) Turaev and Viro${}^{20}$ reformulated the Ponzano--Regge model in
terms of quantum groups providing a combinatorial, topological field
theory in three dimensions with regularization intrinsic within the
structure. This was accomplished by the discovery of a generalization of
the Ponzano--Regge model to the quantum group $U_q(SU(2))$ associated to
$q = e^{2 \pi i / (k + 2)}$ with k being an integer such that the edge
lengths of the diagrams are less than $k/2$. 2.) Crane and
Yetter${}^{21}$ extended the Ponzano--Regge--Turaev--Viro model to four
dimensions using modular tensor categories. 3.) Ooguri${}^{22}$ related
the combinatorial, canonical, and topological approaches for quantum
gravity in three dimensions and developed a realization of a theory in
four dimensions consistent with the Crane--Yetter formalism. 4.)
Rovelli${}^{23}$ found a direct relation between the basis of the
Ponzano--Regge--Turaev--Viro--Ooguri model and the loop representation
basis. This allowed a direct, physical interpretation of the discrete
length spectrum in the Ponzano--Regge--Turaev--Viro--Ooguri model in
terms of the intersection of Wilson lines with the edges. This is
directly related to the discrete areas in loop representation theory due
to the crossing of loops with the surfaces${}^{15}$. This merging of
three of the modern formalisms used to attack the quantum gravity
problem provides further impetus to explore the adaptation of the
Ponzano--Regge model to four dimensions.

We will find that the building blocks for $4$--dimensional quantum
gravity exist within the Ponzano--Regge (and
Ponzano--Regge--Turaev--Viro) model without any kind of extension. A
simple reinterpretation of the model provides a simple link from three to
four dimensions with the proper local Minkowski signature. However, we
must first examine the property that differentiates the spin network
diagrams from geometries---the allowed breaking of the tetrahedral
inequalities. \bigskip

\centerline{\bf IV. PREGEOMETRIC QUANTUM GRAVITY} \medskip

\centerline{\bf A. The Breaking of Tetrahedral Inequalities}
\centerline{\bf and the Origin of Time} \medskip

The primary motivation for considering spin network diagrams is that
they are fundamentally simpler than simplicial geometries, thus the
reference to them as a pregeometry. The tetrahedral inequalities, which
are a fundamental property of tetrahedral--based $3$--geometries, are
violated by the diagrams of some of the spin network configurations,
thus spin networks are less constrained than geometries (more austere).

In their analysis, Hasslacher and Perry${}^{14}$ considered two groups of
configurations of spin network diagrams within the spin sum
($3nj$--symbol), those that violate the tetrahedral inequalities
(somewhere), and those that do not (anywhere). They associated the
configurations which satisfy the tetrahedral inequalities with
$3$--geometries and associated those that violate the tetrahedral
inequalities with topological fluctuations. In this note we will
reinterpret the configurations which violate the tetrahedral
inequalities as the key elements for connecting $3$--geometries into
$4$--geometries.

It should be noted that, unlike in Regge calculus${}^{18}$, we shall not
consider the geometry of the ``interiors" of the tetrahedra of a
diagram. The only ``lengths" will be those associated with edges and the
only ``coordinates" will be those associated at the vertices. Geometric
properties associated with the interiors of the tetrahedra (i.e.
dihedral angles) fall out of the spin sums naturally in the asymptotic
(semiclassical) limit.

Before we return to the Ponzano--Regge model, let us examine the
geometric nature of tetrahedra which violate the tetrahedral
inequalities. We shall consider two distinct inequalities. The first
inequality is evolved from a flat tetrahedron ($2$--dimensional) where
the dihedral angle at one of the edges goes to $\pi$ radians. This
tetrahedron is represented in Fig. 10. There are six proper distance
equations for the five coordinates and the one fully dependent length
($L_6$ is chosen as the dependent length): \begin{equation} {x}_{1}^{\
2}\ =\ {L}_{1}^{\ 2} \label{eq-propdist1} \end{equation} \begin{equation}
{x}_{2}^{\ 2}\ +\ {y}_{2}^{\ 2}\ =\ {L}_{4}^{\ 2} \label{eq-propdist2}
\end{equation} \begin{equation} ({\ x}_{3}\ -{\ x}_{2}{\ )}^{\ 2}\ +\
({\ y}_{3}\ -{\ y}_{2}\ {)}^{\ 2}\ =\ {L}_{3}^{\ 2} \label{eq-propdist3}
\end{equation} \begin{equation} ({\ x}_{3}\ -{\ x}_{1}{\ )}^{\ 2}\ +\
{y}_{3}^{\ 2}\ =\ {L}_{2}^{\ 2} \label{eq-propdist4} \end{equation}
\begin{equation} ({\ x}_{2}\ -{\ x}_{1}{\ )}^{\ 2}\ +\ {y}_{2}^{\ 2}\ =\
{L}_{5}^{\ 2} \label{eq-propdist5} \end{equation} \begin{equation}
{x}_{3}^{\ 2}\ +\ {y}_{3}^{\ 2}\ =\ {L}_{6}^{\ 2} \, ,
\label{eq-propdist6} \end{equation} where expressions
(\ref{eq-propdist1}) through (\ref{eq-propdist5}) are used to solve for
the coordinates in terms of the proper distances $L_i$, $i =
1,\ldots,5$; and expression (\ref{eq-propdist6}) is used to solve for
the proper distance $L_6$.

Solving these equations is difficult and not real enlightening so let us
assume that we have done the solution and have determined $L_6$. Let us
now break one of the tetrahedral inequalities by a small positive amount
$\varepsilon$ by letting $L^{\ 2}_6$  go to $L^{\ 2}_6 + \varepsilon$.
Expression (\ref{eq-propdist6}) is now independent and a new coordinate
is required. We choose to associate this new coordinate ${z}_{3}^{'}$
with vertex B which now has coordinates $({x}_{3}^{'}, {y}_{3}^{'},
{z}_{3}^{'})$. The remaining coordinates are chosen to be fixed so that
expressions (\ref{eq-propdist1}), (\ref{eq-propdist2}), and
(\ref{eq-propdist5}) remain the same. Expressions (\ref{eq-propdist3}),
(\ref{eq-propdist4}), and (\ref{eq-propdist6}) become: \begin{equation}
{{z}_{3}^{'}}^{2}\ +\ ({\ x}_{3}^{'}\ -{\ x}_{2}{\ )}^{\ 2}\ +\ ({\
y}_{3}^{'}\ -{\ y}_{2}\ {)}^{\ 2}\ =\ {L}_{3}^{\ 2} \label{eq-propdist3p}
\end{equation} \begin{equation} {{z}_{3}^{'}}^{2}\ +\ ({\ x}_{3}^{'}\
-{\ x}_{1}{\ )}^{\ 2}\ +\ {{y}_{3}^{'}}^{2}\ =\ {L}_{2}^{\ 2}
\label{eq-propdist4p} \end{equation} \begin{equation} {{z}_{3}^{'}}^{2}\
+\ {{x}_{3}^{'}}^{2}\ +\ {{y}_{3}^{'}}^{2}\ =\ {L}_{6}^{\ 2}\ +\
\varepsilon \, . \label{eq-propdist6p} \end{equation}

Letting ${x}_{3}^{'} = x_3 + \delta x_3$ and ${y}_{3}^{'} = y_3 + \delta
y_3$ in these expressions, and using expressions (\ref{eq-propdist3}),
(\ref{eq-propdist4}), and (\ref{eq-propdist6}) to eliminate terms, we get
\begin{equation} {{z}_{3}^{'}}^{2}\ +\ 2{\ (\ x}_{3}\ -{\ x}_{2}\ )\
\delta {\rm x}_{\rm 3}\rm \ +\ 2\ ({\ y}_{3}\ -{\ y}_{2}\ )\ \delta {\rm
y}_{\rm 3}\rm \ +\ {\delta {\rm x}_{\rm 3}}^{2}\ +\ {\delta {\rm y}_{\rm
3}}^{2}\ =\ 0 \label{eq-propdist3pp} \end{equation} \begin{equation}
{{z}_{3}^{'}}^{2}\ +\ 2{\ (\ x}_{3}\ -{\ x}_{1}\ )\ \delta {\rm x}_{\rm
3}\rm \ +\ 2{\ y}_{3}\ \delta {\rm y}_{\rm 3}\rm \ +\ {\delta {\rm
x}_{\rm 3}}^{2}\ +\ {\delta {\rm y}_{\rm 3}}^{2}\ =\ 0
\label{eq-propdist4pp} \end{equation} \begin{equation}
{{z}_{3}^{'}}^{2}\ +\ 2{\ x}_{3}\ \delta {\rm x}_{\rm 3}\rm \ +\ 2{\
y}_{3}\ \delta {\rm y}_{\rm 3}\rm \ +\ {\delta {\rm x}_{\rm 3}}^{2}\ +\
{\delta {\rm y}_{\rm 3}}^{2}\ =\ \varepsilon \, . \label{eq-propdist6pp}
\end{equation}

Solving these three expressions for $\delta x_3$, and $\delta y_3$, and
${z}_{3}^{'}$, and ignoring higher--order terms in $\varepsilon$, yields
\begin{equation} \delta {\rm x}_{\rm 3}\rm \ =\ {\frac{\varepsilon }{\rm
2\ {x}_{1}}}\rm \ \ \ \ \ \ \ \ \ \ \ \ \ \ (\ >\ 0\ ) \label{eq-delx3}
\end{equation} \begin{equation} \delta {\rm y}_{\rm 3}\rm \ =\
{\frac{\varepsilon \rm \ (\ {x}_{1}\ -\ {x}_{2}\ )}{2\ {x}_{1}\
{y}_{2}}}\rm \ \ \ \ \ \left\{ \begin{array}{1c}>\ 0,\ \ \ {\rm if}\ \
(\ {x}_{1}\ -\ {x}_{2}\ )\ >\ 0\\ <\ 0,\ \ \ {\rm if}\ \ (\ {x}_{1}\ -\
{x}_{2}\ )\ <\ 0\end{array} \right. \label{eq-dely3} \end{equation}
\begin{equation} {z}_{3}^{'}\ =\ \pm \rm \ {\varepsilon }^{1/2}\ \left[
\ 1\ -\ {\frac{{x}_{3}}{{x}_{1}}}\ -\ (\ 1\ -\
{\frac{{x}_{2}}{{x}_{1}}})\ {\frac{{y}_{3}}{{y}_{2}}} {\right]}^{1/2} \,
. \label{eq-z3p} \end{equation}

Note that ${z}_{3}^{'}$ is imaginary if the terms within the second
radical in expression (\ref{eq-z3p}) are negative \begin{equation} 1\ -\
{\frac{{x}_{3}}{{x}_{1}}}\ -\ (\ 1\ -\ {\frac{{x}_{2}}{{x}_{1}}})\
{\frac{{y}_{3}}{{y}_{2}}}\ <\ 0 \, . \end{equation}

Solving this inequality for $y_3$, and using expression (\ref{eq-dely3})
with the definition of ${y}_{3}^{'}$, yields \begin{equation}
{y}_{3}^{'}\ >\ {y}_{3}\ >\ {\frac{{y}_{2}}{{x}_{2}\ -\ {x}_{1}}}\
{x}_{3}\ -\ {\frac{{y}_{2}}{{x}_{2}\ -\ {x}_{1}}}\ {x}_{1}\ \ \ \ \ \ \
\ {\rm if}\ \ (\ {x}_{1}\ -\ {x}_{2}\ )\ >\ 0 \label{eq-ineq11}
\end{equation} or \begin{equation} {y}_{3}^{'}\ <\ {y}_{3}\ <\
{\frac{{y}_{2}}{{x}_{2}\ -\ {x}_{1}}}\ {x}_{3}\ -\
{\frac{{y}_{2}}{{x}_{2}\ -\ {x}_{1}}}\ {x}_{1}\ \ \ \ \ \ \ \ {\rm if}\
\ (\ {x}_{1}\ -\ {x}_{2}\ )\ <\ 0 \, . \label{eq-ineq12} \end{equation}

The expressions on the right of both of these inequalities is the
equation for the line passing through vertices A and C. These two
inequalities are depicted graphically in Fig. 11. It is seen that, for
all points in the regions depicted by the inequalities, our extra
coordinate ${z}_{3}^{'}$ is imaginary. It can thus be concluded that the
tetrahedron becomes timelike upon the breaking of this tetrahedral
inequality.

Let us examine the breaking of the other tetrahedral inequality. This
case evolves from a flat tetrahedron ($2$--dimensional) where the
dihedral angle at one of the edges goes to zero radians. This
tetrahedron is represented in Fig. 12. The expressions
(\ref{eq-propdist1}) through (\ref{eq-propdist6}) still apply as in the
previous case. Let us now break the other tetrahedral inequality by a
small positive amount $\varepsilon$ by letting $L^{\ 2}_6$ go to $L^{\
2}_6 - \varepsilon$. As before, expression (\ref{eq-propdist6}) is now
independent and a new coordinate is required. We choose to associate
this new coordinate ${z}_{3}^{'}$ with vertex B which now has
coordinates $({x}_{3}^{'}, {y}_{3}^{'}, {z}_{3}^{'})$. The remaining
coordinates are chosen to be fixed so that equations
(\ref{eq-propdist1}), (\ref{eq-propdist2}), and (\ref{eq-propdist5})
remain the same. Expressions (\ref{eq-propdist3}), (\ref{eq-propdist4}),
and (\ref{eq-propdist6}) become \begin{equation} {{z}_{3}^{'}}^{2}\ +\
({\ x}_{3}^{'}\ -{\ x}_{2}{\ )}^{\ 2}\ +\ ({\ y}_{3}^{'}\ -{\ y}_{2}\
{)}^{\ 2}\ =\ {L}_{3}^{\ 2} \label{eq-propdist3prp} \end{equation}
\begin{equation} {{z}_{3}^{'}}^{2}\ +\ ({\ x}_{3}^{'}\ -{\ x}_{1}{\
)}^{\ 2}\ +\ {{y}_{3}^{'}}^{2}\ =\ {L}_{2}^{\ 2} \label{eq-propdist4prp}
\end{equation} \begin{equation} {{z}_{3}^{'}}^{2}\ +\ {{x}_{3}^{'}}^{2}\
+\ {{y}_{3}^{'}}^{2}\ =\ {L}_{6}^{\ 2}\ -\ \varepsilon \, ,
\label{eq-propdist6prp} \end{equation} respectively.

Letting ${x}_{3}^{'} = x_3 + \delta x_3$ and ${y}_{3}^{'} = y_3 + \delta
y_3$ in these expressions, and using expressions (\ref{eq-propdist3}),
(\ref{eq-propdist4}), and (\ref{eq-propdist6}) to eliminate terms, we get
\begin{equation} {{z}_{3}^{'}}^{2}\ +\ 2{\ (\ x}_{3}\ -{\ x}_{2}\ )\
\delta {\rm x}_{\rm 3}\rm \ +\ 2\ ({\ y}_{3}\ -{\ y}_{2}\ )\ \delta {\rm
y}_{\rm 3}\rm \ +\ {\delta {\rm x}_{\rm 3}}^{2}\ +\ {\delta {\rm y}_{\rm
3}}^{2}\ =\ 0 \label{eq-propdist3prpp} \end{equation} \begin{equation}
{{z}_{3}^{'}}^{2}\ +\ 2{\ (\ x}_{3}\ -{\ x}_{1}\ )\ \delta {\rm x}_{\rm
3}\rm \ +\ 2{\ y}_{3}\ \delta {\rm y}_{\rm 3}\rm \ +\ {\delta {\rm
x}_{\rm 3}}^{2}\ +\ {\delta {\rm y}_{\rm 3}}^{2}\ =\ 0
\label{eq-propdist4prpp} \end{equation} \begin{equation}
{{z}_{3}^{'}}^{2}\ +\ 2{\ x}_{3}\ \delta {\rm x}_{\rm 3}\rm \ +\ 2{\
y}_{3}\ \delta {\rm y}_{\rm 3}\rm \ +\ {\delta {\rm x}_{\rm 3}}^{2}\ +\
{\delta {\rm y}_{\rm 3}}^{2}\ =\ -\ \varepsilon \, .
\label{eq-propdist6prpp} \end{equation}

Solving these three expressions for $\delta x_3$, and $\delta y_3$, and
${z}_{3}^{'}$, and ignoring higher--order terms in $\varepsilon$, yields
\begin{equation} \delta {\rm x}_{\rm 3}\rm \ =\ \rm -\
{\frac{\varepsilon }{\rm 2\ {x}_{1}}}\ \ \ \ \ \ \ \ \ \ \ \ \ \ (\ <\
0\ ) \label{eq-deltx3} \end{equation} \begin{equation} \delta {\rm
y}_{\rm 3}\rm \ =\ {\frac{\varepsilon \rm \ (\ {x}_{2}\ -\ {x}_{1}\
)}{2\ {x}_{1}\ {y}_{2}}}\rm \ \ \ \ \ \ \ \ \ \left\{
\begin{array}{1c}>\ 0,\ \ \ {\rm if}\ \ (\ {x}_{2}\ -\ {x}_{1}\ )\ >\
0\\ <\ 0,\ \ \ {\rm if}\ \ (\ {x}_{2}\ -\ {x}_{1}\ )\ <\ 0\end{array}
\right. \label{eq-delty3} \end{equation} \begin{equation} {z}_{3}^{'}\
=\ \pm \rm \ {\varepsilon }^{1/2}\ \left[ \ -\ 1\ +\
{\frac{{x}_{3}}{{x}_{1}}}\ +\ (\ 1\ -\ {\frac{{x}_{2}}{{x}_{1}}})\
{\frac{{y}_{3}}{{y}_{2}}}{\right]}^{1/2} \, . \label{eq-z3pr}
\end{equation}

Note that ${z}_{3}^{'}$ is imaginary if the terms within the second
radical in equation (18) are negative \begin{equation} -\ 1\ +\
{\frac{{x}_{3}}{{x}_{1}}}\ +\ (\ 1\ -\ {\frac{{x}_{2}}{{x}_{1}}})\
{\frac{{y}_{3}}{{y}_{2}}}\ <\ 0 \, . \end{equation}

Solving this inequality for $y_3$, and using expression
(\ref{eq-delty3}) with the definition of ${y}_{3}^{'}$, yields
\begin{equation} {y}_{3}^{'}\ >\ {y}_{3}\ >\ {\frac{{y}_{2}}{{x}_{2}\ -\
{x}_{1}}}\ {x}_{3}\ -\ {\frac{{y}_{2}}{{x}_{2}\ -\ {x}_{1}}}\ {x}_{1}\ \
\ \ \ \ \ \ {\rm if}\ \ (\ {x}_{2}\ -\ {x}_{1}\ )\ >\ 0 \label{eq-ineq21}
\end{equation} or \begin{equation} {y}_{3}^{'}\ <\ {y}_{3}\ <\
{\frac{{y}_{2}}{{x}_{2}\ -\ {x}_{1}}}\ {x}_{3}\ -\
{\frac{{y}_{2}}{{x}_{2}\ -\ {x}_{1}}}\ {x}_{1}\ \ \ \ \ \ \ \ {\rm if}\
\ (\ {x}_{2}\ -\ {x}_{1}\ )\ <\ 0 \, . \label{eq-ineq22} \end{equation}

Again, the expressions on the right of both of these inequalities is the
equation for the line passing through vertices A and C. These two
inequalities are depicted graphically in Fig. 13. It is seen that, for
all points in the regions depicted by the inequalities, our extra
coordinate ${z}_{3}^{'}$  is imaginary. It can thus be concluded that
the tetrahedron becomes timelike upon the breaking of this tetrahedral
inequality also.

This means that a tetrahedron in a Euclidean $3$--geometry which breaks
the tetrahedral inequalities will acquire a timelike nature (a timelike
region on the spacelike hypersurface) and define a local time scale (the
${z}_{3}^{'}$ in our example tetrahedra above). The concept of time is
derived from allowing the breaking of tetrahedral inequalities---a
relaxation of a natural property of geometries. Thus a region of a given
tetrahedral spacelike hypersurface may become timelike by breaking the
tetrahedral inequalities in this region.

However, we can achieve more than this from the spin network diagrams.
We can show that some spin network configuration diagrams may be
associated with actual slices of locally--Minkowski $4$--geometry.
\bigskip

\centerline{\bf B. Constructing a Slice of $4$--Geometry}
\centerline{\bf from Spin Network Diagrams} \medskip

Consider the following construction. Start with a spin configuration
diagram whose connectivity is that of a tetrahedral $3$--geometry with
the vertices labelled with $T_i$. We shall use two spin identities in
our construction which allow the tetrahedra of the diagram to be
subdivided without affecting the spin sum:

\begin{itemize} \item Step 1: Place a vertex $P$ within the ``interior"
of each tetrahedron ($6j$--symbol diagram). Connect each new vertex to
the four vertices $T_i$ of the tetrahedron they are ``contained in"
using four new edges. This subdivides each of the original tetrahedra
into four tetrahedra (see Fig. 14). This construction on each
tetrahedron is the diagrammatic representation of the identity presented
previously in expression (\ref{eq-sixjidentren}) which allows each
$6j$--symbol to be written as the sum of products of four $6j$--symbols.
This construction does not change the overall spin sum ($3nj$--symbol of
the whole lattice).

\item Step 2: Consider each neighboring pair of our new tetrahedra of
Step 1 with a triangle of our original tetrahedral lattice shared
between them. For each pair, connect the $P$ vertices (those added in
Step 1) of the two tetrahedra by a new edge. This subdivides the two
tetrahedra into three tetrahedra (see Fig. 15). This construction on
each of these pair of tetrahedra is the diagrammatic representation of
the identity presented previously in expression (\ref{eq-beidentity})
which allows each pair of $6j$--symbols to be written as the sum of
products of three $6j$--symbols (Elliot--Biedenharn identity). Again,
the construction does not change the overall spin sum. Note that the set
of the new edges added in this step form a lattice which is
geometrically dual to the original tetrahedral lattice. The blocks of
this dual lattice are various polyhedra whose form depend on the
connectivity of the original tetrahedral lattice.

\item Step 3: Consider each polygon face of the dual cells formed by the
edges of Step 2. Each polygon is shared by two dual lattice polyhedra
each of which ``contain" a vertex $T$ ($T'$) of the original tetrahedral
lattice in their ``interior". There is a single edge of the original
tetrahedral lattice ``piercing" the plane of the given polygon
(connecting the two vertices $T$ and $T'$). Furthermore, $T$ and $T'$
are connected by edges from Step 1 to each of the vertices $P_i$ of the
given polygon. This group contains $n$ tetrahedra sharing (in the
entourage of) the edge between $T$ and $T'$, where $n$ is the number of
edges of the given polygon.

Now, place new edges between the vertices of the given polygon such that
the polygon is subdivided into triangles (there may be more than one way
to do this but it does not matter to the construction) (see Fig. 16).
This subdivides the original $n$ tetrahedra into $2n-3$ tetrahedra ($3$
tetrahedra around the $T$--$T'$ edge associated with one of the new
triangles and $2$ tetrahedra sharing each of the remaining triangles).
This step is just the repeated application of the same identity as in
Step 2 but with the new edges piercing triangles which are not part of
the original tetrahedral lattice. \end{itemize}

Performing Step 3 on every polygon face yields a tetrahedral lattice
$3$--geometry with the same connectivity as a fully--rigidified
null--strut sandwich of $4$--geometry${}^{24}$ (see Fig. 17). This is
the key observation. If we allow the new edges of Step 1 to violate the
tetrahedral inequalities, which is allowed for spin configurations, the
remaining new edges are forced out of the original $3$--geometry by the
fact that one or more of the four tetrahedra per original tetrahedron of
step 1 become timelike. This new lattice configuration is a slice of
$4$--geometry with exactly the same connectivity as a null--strut
lattice. The original tetrahedral lattice forms one of the bounding
$3$--geometries which we shall call the $TET$ hypersurface. The lattice
formed by the rigidified (triangulated) dual cells is the other bounding
$3$--geometry which we shall call the $TET^{\ast}$ hypersurface. This
lattice has been found to be a dynamically--consistent slice of
$4$--geometry in the classical Regge calculus formalism. It has the
necessary degrees of freedom, and an equal number of Regge equations and
$TET^{\ast}$ edge length variables consistent with a classical $3+1$
evolution procedure.

A true null--strut lattice cannot be achieved by these spin lattice
configurations because that would require that the triangle inequalities
be violated and that zero lengths be allowed (for struts to become
null). Neither of these is allowed for spin networks. However, as has
been stated, a lattice with this connectivity has been found to be a
dynamically--consistent slice of $4$--geometry. This consistency is not
dependent on the existence of null--struts in the lattice, but is
dependent on the connectivity of the lattice and the simple properties
of $4$--dimensional polytopes${}^{25}$. Furthermore, the null--strut
lattice may be ``approached" as we take the continuum limit.

Some other profound observations can be made concerning this
construction of $4$--geometry. 1.) The null--strut lattice is compelling
because it is built from a combination of blocks which are lattice
analogues of light cones (wigwam simplices and fluted cones) and filler
blocks which are contained entirely within the sandwich (no $3$--blocks
from these fillers are within the bounding $3$--geometries), thus
indicating the fundamental nature of light cones within this pregeometry
formalism as the continuum limit is approached. Indeed, attempts to
construct lattice sandwiches of $4$--geometry with other connectivities
using the spin identities have failed (prism--type lattices for
example${}^{26}$). It should also be stressed that the fluted cones must
be treated as light cones (interior of fluted cone is assumed to be flat
Minkowski --- no interior edges necessary for rigidification) in order
to have an equal number of Regge equations and $TET^{\ast}$ edge length
variables for classical evolution. {\it This implies that local
Minkowski structure is necessary for the consistency of classical
dynamics.} 2.) The second--order nature of the dynamics of
$4$--geometries is implied by the fact that there is a two--valued
degeneracy in the breaking of the tetrahedral inequalities based on the
choice of the sign of the new coordinate. 3.)  The spin identities have
been employed to achieve $3$--dimensional structure from the
$2$--dimensional spin diagrams and the breaking of the tetrahedral
inequalities have been used to achieve $4$--dimensional,
locally--Minkowski sandwich structures from the $3$--dimensional
structures (see Fig. 18). This strongly implies the origin of spacetime
dimension.

It can be concluded that the set of spin network configuration diagrams
contain both the set of all closed $3$--geometries (spacelike, timelike,
and spacelike with timelike regions) and the set of possible
$4$--dimensional sandwich structures which connect them. Therefore, this
set contains all the building blocks necessary for the construction of
$4$--dimensional spacetime. Our new model for pregeometry is interpeted
as a utilization of the PCM in Fig. 19 based on the fundamental
constants $\hbar$ (fundamental length) and $c$ (ratio between
fundamental length and fundamental time). \bigskip

\centerline{\bf C. The Semiclassical and Classical Limit Implications}
\medskip

This new building block for spacetime structure has some important
implications to the Hartle--Hawking semiclassical path integral in four
dimensions${}^4$:

\begin{itemize} \item If we consider the product inside the integral of
expression (\ref{eq-3dpathint}) to provide a physically--derived measure
for the $3$--geometries within the semiclassical path integral in four
dimensions (this has already been observed in the Ooguri model), the
integral is dominated by those spin network configurations that satisfy
the tetrahedral inequalities. This is due to the fact that the weight of
the configurations that violate the tetrahedral inequalities approach
zero relative to the measures of some of the configurations which
satisfy the tetrahedral inequalities as the semiclassical limit is
approached. Therefore, the $3$--geometries remain as the basis for the
phase space for $4$--dimensional quantum gravity. The configurations
that violate the tetrahedral inequalities act only to provide a basis
for the relating of $3$--geometries into $4$--geometries (extrinsic
curvature information).

\item The Hartle--Hawking path integral in four dimensions now has a
physically--derived renormalization cut--off: the smallest possible spin
value   and a naturally--defined renormalization scale parameter
associated with the fineness of the tesselation of a given spin network
($p$). Furthermore, the discreteness of the lengths (spins), a property
which remains at the quantum level, breaks the exact conformal
invariance, thus solving the positive--definiteness problem. This
discreteness also prohibits zero--volume $4$--blocks, thus suppressing
the formation of singularities. Therefore, the three most bothersome
problems in quantum gravity are avoided.

\item The spin network building blocks of this model provide a structure
which can include various topologies in the $4$--dimensional
configurations. They can be constructed by connecting sandwich
configurations with isolated timelike regions together at these regions
leaving handles in the $4$--geometry. \end{itemize}

The question of the origin of the Einstein equations (and the
Einstein--Hilbert action) in the classical limit remains. The work of
Ooguri on $4$--dimensional topological lattice models associated with
the group $SU(2)$ introduced an action in analogy with $\lambda \phi^4$
theory with an associated partition function which was shown to be a
topological invariant for $4$-manifolds (invariant under the Alexander
moves${}^{27}$), much as the lattice in Ponzano--Regge model is a
topological invariant for $3$--manifolds. This action is compelling, and
may indeed describe the quantum theory, but the connection with the
Einstein--Hilbert action and the origin of its functional form from
first principles is unclear at present. In contrast, the connection and
form in the Ponzano--Regge model becomes clear upon taking the
semiclassical limit. Furthermore, the Ooguri model relies on the
assumption of $4$--dimensional structure in contrast to the model
presented here which derives four dimensions as a result of the
properties of the spin structures. Note that we do not consider the
$4$--dimensional model presented here to be a purely topological field
theory (as in the Ooguri model). The origin of the Einstein--Hilbert
action may be described in an entirely simple (austere) fashion as
follows.

In the classical limit, a $4$--geometry, built from the discrete
$4$--geometry--like sandwiches of the quantum regime, becomes a manifold
effectively (the discrete nature is at a much smaller scale than any
local radii of curvature). The diffeomorphism invariance, which is only
an approximate symmetry on discrete lattices, becomes an exact symmetry
at the classical limit${}^{24}$. The key to acquiring the dynamic
equations of the $3$--geometries embedded in our $4$--geometry lies in
the work of Hojman, Kuch\v{a}r, and Teitelboim${}^{28}$. They found that
the dynamics of $3$--geometry is already encoded in the structure of the
$4$--geometry.

Consider the classical evolution of a $3$--geometry from a given
hypersurface $\sigma_1$ to another hypersurface $\sigma_2$ (see Fig.
20). The form of the hypersurface $\sigma_2$ obtained by the evolution
procedure must be independent of the choice of foliation used in the
evolution between the two hypersurfaces (dotted slices versus dashed
slices in Fig. 20).  The only second--order dynamic equations that leads
to dynamics of a given $3$--geometry which is independent of the choice
of time slicing used in the evolution is the Einstein equations (with
possible cosmological term).

The arguement is fairly simple. Consider the dynamical law for the
change in a functional $F$ of the canonical variables $\phi^A$, $\pi_B$
of a hypersurface under a deformation of the hypersurface $\delta \xi$
\begin{equation} \delta F\ = \ \int_{}^{}dx\ \{[ F\ ,\ {H}_{\perp }(x)]\
\delta {\xi }^{\perp } (x)\ +\ [ F\ ,\ {H}_{i}(x)]\ \delta {\xi }^{ i}
(x)\}\ \equiv \ \int_{}^{}dx\ [ F\ ,\ {H}_{\mu }]\ \delta {\xi }^{\mu }
\, , \label{eq-delF} \end{equation} where ${H}_{\perp }$ is the
super--Hamiltonian and ${H}_{i}$ is the supermomentum. It was shown by
Hojman et. al., using repeated applications of expression
(\ref{eq-delF}), that for the change in $F$ from $\sigma_1$ to
$\sigma_2$ to be independent of the series of infinitesimal deformations
used to go from one to the other (independent of the choice of time
slicing), five conditions must be satisfied. The three Poisson bracket
closure relations \begin{equation} [{H}_{\alpha }(x)\ ,\ {H}_{\beta
}(x')]\ =\ \int_{}^{}dx''\ {\kappa }_{\alpha \beta }^{\ \ \ \gamma
}(x'';\ x\ ,\ x')\ {H}_{\gamma }(x'') \, , \label{eq-closure}
\end{equation} where \begin{equation} {\kappa }_{r\perp }^{\ \ \  \perp
}(x'';\ x\ ,\ x')\ =\ -{\kappa }_{\perp r}^{\ \ \  \perp }(x'';\ x'\ ,\
x)\ =\ \delta (x'',\ x)\ {\delta }_{,r}(x'',\ x') \, , \label{eq-kap1}
\end{equation} \begin{equation} \begin{array}{1c}{\kappa }_{mn}^{\ \ \
r}(x'';\ x\ ,\ x')\ =\ -{\kappa }_{nm}^{\ \ \ r}(x'';\ x'\ ,\ x)\\ \ \ \
\ \ \ \ \ \ \ \ \ \ \ \ \ \ \ \ \ \ \ \ =\ \delta (x'',\ x)\ {\delta
}_{,m}(x'',\ x'){\delta }_{n}^{r}\ -\ \delta (x'',\ x')\ {\delta
}_{,n}(x'',\ x){\delta }_{m}^{r} \, ,\end{array} \label{eq-kap2}
\end{equation} and \begin{equation} \begin{array}{1c}{\kappa }_{\perp
\perp }^{\ \ \ r}(x'';\ x\ ,\ x')\ =\ -{\kappa }_{\perp \perp }^{\ \ \
r}(x'';\ x'\ ,\ x)\\ \ \ \ \ \ \ \ \ \ \ \ \ \ \ \ \ \ \ \ \ \ \ \ =\
\varepsilon \ {g}^{rs}(x'')\ [\delta (x'',\ x')\ {\delta }_{,s}(x'',\
x)\ -\ \delta (x'',\ x)\ {\delta }_{,s}(x'',\ x')] \, ,\end{array}
\label{eq-kap3} \end{equation} (obtained from geometrical arguements
about the general inequivalence of commuting deformations) must be
satisfied, where variable $\varepsilon$ is the spacetime signature
(known to be -1 from our spin network derived $4$--geometry) and
${g}^{rs}$ is the spatial metric on the hypersurface. Furthermore, the
two constraint equations, \begin{equation} {H}_{\perp }\ =\ 0
\label{eq-constr1} \end{equation} and \begin{equation} {H}_{i}\ =\ 0
\label{eq-constr2} \end{equation} must also be satisfied. Furthermore,
it was shown, in the case where the canonical variables are the spatial
metric $g_{ij}$ and its conjugate momentum ${\pi}^{kl}$, that the
solution of the Poisson bracket closure relations uniquely determine the
form of the super--Hamiltonian and the supermomentum, \begin{equation}
{H}_{\perp }^{grav}\ =\ \kappa {g}^{-1/2} ({g}_{ik}{g}_{jl}\ +\
{g}_{il}{g}_{jk}\ -\ {g}_{ij}{g}_{kl}){\pi }^{ij}{\pi }^{kl}\ +\
\varepsilon {(2\kappa)}^{-1}{g}^{1/2} (R\ -\ 2\lambda) \label{eq-supham}
\end{equation} and \begin{equation} {H}_{i}^{grav}\ =\ -2{\pi }_{i\ \
|s}^{\ s} \, \, , \label{eq-supmom} \end{equation} where R is the Ricci
scalar of the $4$--metric and the $|$ in expression (\ref{eq-supmom})
denotes a derivative with respect to the spatial coordinates. These are
precisely the generators for Einstein's equations with a cosmological
constant $\lambda$.

In other words, the dynamic and constraint equations are uniquely
determined by the utterly simple requirement of consistency
(embeddability requirement) of the resulting classical dynamics. This
``dynamics self--encoded within the structure evolved" concept is
extremely powerful and general. Applying the same embeddability
criterion to vector fields and second--order equations leads to the
Maxwell's equations and similar application to vector fields with
internal spin degrees of freedom leads to the Yang--Mills theory for
quark binding.

To add to the strength of this self--encoding arguement, it should also
be noted that the conservation laws (and therefore the constraints) for
all three of these theories are self--encoded within the background
$4$--geometry through a simple, tautological, property of topology, the
``boundary of a boundary" principle, as has been noted by Kheyfets and
Miller${}^{29}$. Application of the principle in its
$2,3,4$--dimensional form to $4$--metrics yields the contracted Bianchi
Identities \begin{equation} \nabla \rm \ T\ =\ \nabla \rm \ \nabla \rm \
\Gamma \rm \ =\ 0 \, , \label{eq-bianchiid} \end{equation} where $T$ is
the energy-momentum tensor and $\Gamma$ is the connection. Similar
application of the principle to electromagnetic fields lead to the
current conservation law \begin{equation} d\ J\ =\ {d}^{2}\ {}^*F\ =\ 0
\, , \label{eq-emconserv} \end{equation} where $J$ is the $4$--current
and ${}^*F$ is the dual of the Faraday tensor. Finally, application to
Yang--Mills fields leads to the conservation of field source current
\begin{equation} D\ J\ =\ D\ D\ {}^*B\ =\ 0 \, , \label{eq-ymconserv}
\end{equation} where $J$ is the field source current and $B$ is the
Yang-Mills field.

Therefore, in summary, once the $4$--geometries are formed (in the
quantum regime), they already have their classical dynamics encoded in
their structure---they ``know" how to evolve classically. No additional
primordial encoding within the spin structures is necessary.
Furthermore, the embeddability criterion and the ``boundary of a
boundary" principle form extremely simple unification principles for the
great field theories of physics. The conjecture by Wheeler suggesting
austerity as a unifying principle appears to be correct. \bigskip

\centerline{\bf V. FUTURE CONSIDERATIONS} \medskip

Taking the pregeometric philosophy a step further, two extensions of
this work are being pursued. The first involves a relaxation of the spin
addition laws to allow for the breaking of the triangle inequalities as
well as the tetrahedral inequalities. This is being done by analytically
continuing the definitions of Wigner $3j$--symbols (expression
(\ref{eq-3jsymbol})), replacing the factorials with Gamma functions.

The second extension involves recent work by Kheyfets, LaFave, Miller,
and  Whooters${}^{30}$ to formulate simplicial geometries from
information networks. These extremely austere networks are described by
$N$--dimensional simplicial diagrams with s-state random variables at
the nodes (verticies). The simplex, described by the set of pairwise
probabilities for the data (given along the edges), is determined by
simple constraints on the pairwise probabilities as well as the
information distance given along the edges. Remarkably, it was found
that single binary variables at the nodes of the information simplex
were unique in the obtaining of equation/variable consistency (fully
determines the lattice ``geometry"), regardless of the dimension $N$.
However, there is nothing in the formalism, as of yet, which explains
the source of classical dimensionality as is true of the spin networks
in this paper. To alleviate this shortcoming, an attempt to describe the
spin networks as a type of information network is being pursued in order
to bridge the information networks to the $4$--dimensional,
locally--Minkowski geometries of classical dynamics. \bigskip

\centerline{ACKNOWLEDGEMENTS} \medskip

        The Author would like to thank the following people for useful
discussions and suggestions: R. Matzner, J. A. Wheeler, P. Laguna, J.
York, W. Miller, A. Kheyfets, W. McCarter, and T. Wilson. \bigskip

\begin{description} \item[${}^1$] J. A. Wheeler, {\it Geometrodynamics},
(Academic Press, New York, 1962); J. A. Wheeler in {\it Battelle
Rencontres: 1967 Lectures in Mathematics and Physics}, edited by C.
DeWitt and J. A. Wheeler (Benjamin, New York, 1968).

\item[${}^2$] A. Lichernowicz, Acad. Sci. Paris, Comptes Rend. ${\bf
252}$, 3742 (1961); A. Lichernowicz, Acad. Sci. Paris, Comptes Rend.
${\bf 253}$, 940 (1961); A. Lichernowicz, Acad. Sci. Paris, Comptes Rend.
${\bf 253}$, 983 (1961); J. Milnor, Enseignment Math. ${\bf 9}$, 198
(1963); J. Milnor, Topology ${\bf 3}$, 223 (1965); J. Milnor in {\it
Cairns 1965} (1965), p. 55; W. -C. Hsiang and B. J. Sanderson, Illinois
J. Math. ${\bf 9}$, 651 (1965); D. W. Anderson, E. H. Brown, Jr., and F.
P. Peterson, Bull. Am. Math. Soc. ${\bf 72}$, 256 (1966); D. W.
Anderson, E. H. Brown, Jr., and F. P. Peterson, Ann. Math. ${\bf 83}$,
54 (1966).

\item[${}^3$] G. F. B. Riemann in {\it Mathematische Werke} (Teubner,
Liepzig, 1876).

\item[${}^4$] J. B. Hartle and S. W. Hawking, Phys. Rev. ${\bf D28}$,
2960 (1983).

\item[${}^5$] W. Haken in {\it Word Problems}, edited by W. Boone, F. B.
Cannonito, and R. C. Lyndon (North--Holland, Amsterdam, 1973).

\item[${}^6$] J. B. Hartle and S. W. Hawking, Phys. Rev. ${\bf D13}$,
2188 (1976); S. W. Hawking, Phys. Lett. ${\bf 60A}$, 81 (1977); G. W.
Gibbons and S. W. Hawking, Phys. Rev. ${\bf D15}$, 2752 (1977).

\item[${}^7$] S. W. Hawking in {\it General Relativity, An Einstein
Centenary Survey}, edited by S. W. Hawking and W. Israel (Cambridge,
1979).

\item[${}^8$] S. Deser in {\it Proceedings of the Conference on Gauge
Theories and Modern Field Theory}, edited by R. Arnowitt and P. Nath
(MIT Press, Cambridge, Massachusetts, 1975); K. S. Stelle, Phys. Rev.
${\bf D16}$, 953 (1977).

\item[${}^9$] G. 't Hooft, Nucl. Phys. ${\bf B62}$, 444 (1973); G. 't
Hooft and M. Veltman, Ann. Inst. Poincar\'{e} ${\bf 20}$, 69 (1974); S.
Deser and P. van Nieuwenhuizen, Phys. Lett. ${\bf 32}$, 245 (1974); S.
Deser and P. van Nieuwenhuizen, Phys. Rev. ${\bf D10}$, 401, 411 (1974);
S. Deser, H--S Tsao, and P. van Nieuwenhuizen, Phys. Lett. ${\bf 50B}$,
491 (1974).

\item[${}^{10}$] L. Smolin in {\it Proceedings of the Osgood Hill
Conference on Conceptual Problems of Quantum Gravity}, edited by A.
Ashtekar and J. Stachel (Birkhauser, 1989).

\item[${}^{11}$] J. A. Wheeler in {\it Quantum Theory and Gravitation},
edited by A. R. Marlow (Academic,1980); J. A. Wheeler in {\it Structure
in Science and Art}, edited by P.Medawar and J. Shelley (North--Holland,
New York, 1980); J. A. Wheeler in {\it Discovery: Research and
Scholarship at the University of Texas at Austin}, ${\bf 7}$(2), 4
(1982); J. A. Wheeler, Amer. J. Phys. ${\bf 51}$, 398 (1983); J. A.
Wheeler in {\it Experimental Gravitation and Measurement Theory}, edited
by P. Meystre and M. Scully (Plenum, New York and London, 1983); J. A.
Wheeler in {\it Problems in Theoretical Physics}, edited by A.
Giovannini, F. Mancini, and M. Marinaro (University of Salerno Press,
Salerno, 1984); J. A. Wheeler, J. of Res. and Dev. ${\bf 32}$, 4 (1988);
J. A. Wheeler, {\it It from Bit: Information, Physics, Quantum, The
Search for Links}, preprint (1990).

\item[${}^{12}$] G. Ponzano and T. Regge in {\it Spectroscopy and Group
theoretical methods in Physics}, edited by F. Bloch (North Holland,
Amsterdam 1968).

\item[${}^{13}$] R. Penrose in {\it Quantum Gravity: an Oxford
Symposium}, edited by E. T. Bastin (Cambridge U. P., 1970); R. Penrose
in {\it Magic without Magic}, edited by J. R. Klauder (Freeman, San
Francisco, 1972); R. Penrose in {\it Quantum Gravity: an Oxford
Symposium} (Oxford U. P., 1975).

\item[${}^{14}$] B. Hasslacher and M. J. Perry, Phys. Lett. ${\bf B
103}$, 21 (1981); S. M. Lewis, Phys. Lett. ${\bf B122}$, 265 (1983).

\item[${}^{15}$] L. Smolin in {\it Proceedings of the 1991 GIFT
International Seminar on Theoretical Physics: Quantum Gravity and
Cosmology} (World Scientific, Singapore, 1992).

\item[${}^{16}$] A. P. Yutsis, I. B. Levinson, and V. V. Vanagas, {\it
Mathematical Apparatus of the Theory of Angular Momentum} (Israel
Program for Scientific Translation, Jerusalem, 1962).

\item[${}^{17}$] L. C. Biedenharn, J. Math. Phys. ${\bf 31}$, 287 (1953);
J. P. Elliot, Proc. Royal Soc. ${\bf A218}$, 370 (1953).

\item[${}^{18}$] T. Regge, Nuovo Cimento ${\bf 19}$, 558 (1961).

\item[${}^{19}$] L. P. Kadanoff, Rev. Mod. Phys. ${\bf 49}$, 267 (1977).

\item[${}^{20}$] V. Turaev and O. Viro, Topology ${\bf 31}$, 865 (1992).

\item[${}^{21}$] L. Crane and D. Yetter, {\it A categorical construction
of 4D topological quantum field theories} (Kansas State University
preprint 1992); L. Crane, {\it Categorical Physics} (Kansas State
University preprint 1992).

\item[${}^{22}$] H. Ooguri and N. Sasakura, Mod. Phys. Lett. ${\bf A6}$,
3591 (1991); H. Ooguri, Nucl. Phys. ${\bf B382}$, 276 (1992); H. Ooguri,
Mod. Phys. Lett. ${\bf A7}$, 2799 (1992).

\item[${}^{23}$] C. Rovelli, {\it The Basis of the
Ponzano--Regge--Turaev--Viro--Ooguri Quantum--Gravity Model is the Loop
Representation Basis} (University of Pittsburgh preprint 1993).

\item[${}^{24}$] W. A. Miller, {\it Foundations of Null--Strut
Geometrodynamics}, Dissertation (The University of Texas, 1986); A.
Kheyfets, N. J. LaFave, and W. A. Miller, Phys. Rev. ${\bf D39}$ (4),
1097 (1988); N. J. LaFave, {\it Simplicial Lattices in Classical and
Quantum Gravity: Mathematical Structure and Application}, Dissertation
(The University of Texas, 1989); A. Kheyfets, N. J. LaFave, and W. A.
Miller, Class. Quantum Grav. ${\bf 6}$, 659 (1989); A. Kheyfets, N. J.
LaFave, and W. A.  Miller, Phys. Rev. ${\bf D41}$, 3628 (1990); A.
Kheyfets, N. J. LaFave, and W. A.  Miller, Phys. Rev. ${\bf D41}$, 3637
(1990).

\item[${}^{25}$] H. S. M. Coxeter, {\it Regular Polytopes} (Dover, New
York, 1973).

\item[${}^{26}$] C. Y. Wong, J. Math. Phys. ${\bf 12}$, 70 (1971); P. A.
Collins and R. M. Williams, Phys. Rev. ${\bf D5}$, 1908 (1972); P. A.
Collins and R. M. Williams, Phys. Rev. ${\bf D7}$, 965 (1973); P. A.
Collins and R. M. Williams, Phys. Rev. ${\bf D10}$, 3537 (1974); J.
Porter, Class. Quantum Grav. ${\bf 4}$, 375 (1987); J. Porter, Class.
Quantum Grav. ${\bf 4}$, 391 (1987); L. Brewin, Class. Quantum Grav.
${\bf 5}$, 839 (1988).

\item[${}^{27}$] J. W. Alexander, Ann. Math. ${\bf 31}$, 292 (1930); M.
Gross and S. Varsted, preprint (1991).

\item[${}^{28}$] S. Hojman, K. Kuch\v{a}r, and C. Teitelboim, Nature
Phys. Sci. ${\bf 245}$, 97 (1973); C. Teitelboim, Ann. Phys. ${\bf 79}$,
542 (1973); C. Teitelboim, {\it The Hamiltonian Structure of Spacetime},
Dissertation (Princeton University, 1973); K. Kuch\v{a}r in {\it
Relativity, Astrophysics, and Cosmology}, edited by W. Israel (Reidel,
Dordrecht, Holland, 1973); K. Kuch\v{a}r, J. Math. Phys. ${\bf 15}$, 708
(1974); C. Teitelboim in {\it Relativity, Fields, Strings, and Gravity:
Proceedings of the Second Latin American Symposium on Relativity and
Gravitation SILARG II}, edited by C. Aragone (Universidad Simon Bolivar,
Caracus, 1976); S. A. Hojman, K. Kuch\v{a}r, and C. Teitelboim, Ann.
Phys. ${\bf 76}$, 88 (1976).

\item[${}^{29}$] A. Kheyfets and W. A. Miller, J. Math. Phys. ${\bf
32}$, 3168 (1991).

\item[${}^{30}$] A. Kheyfets, N. J. LaFave, W. A. Miller, and W.
Whooters, {\it An Information Theoretic Approach to Pregeometry}, paper
in preparation (1993).

\eject

Fig. 1. The ``Meaning Circuit" describing the ``world" as an
information--derived construct defined by observer--participancy. \vfill
\eject

Fig. 2. The meaning circuit ``picks" physics from the set of all
consistent mathematical structures via the participancy of the observers.
\vfill \eject

Fig. 3. The Pregeometry Construct Methodology (PCM). This is an approach
to defining a consistent pregeometry. The methodology verifies the
consistency of the candidate pregeometry and illuminates its properties
by taking the semiclassical limit of the structure and comparing the
resulting structure and expressions with known semiclassical structures
and expressions. \vfill \eject

Fig. 4. The spin diagram of a Wigner $3j$--symbol. The solid lines
represent the diagram we use in our model which is dual to the diagrams
used by Yutsis et al.${}^{}$ (dotted lines). \vfill \eject

Fig. 5. The spin diagram for a $6j$--symbol. \vfill \eject

Fig. 6. A spin diagram of the identity which equates a $6j$--symbol with
a sum of products of four $6j$--symbols. \vfill \eject

Fig. 7. A spin diagram of the Elliot--Biedenharn identity which equates a
product of two $6j$--symbols with a sum of products of three
$6j$--symbols. \vfill \eject

Fig. 8. The spin diagram of the $6j$--symbol used to explain Cayley's
determinant. $j_{kl}$ is the ``length" of an edge in terms of the spin
value $a_{kl}$ \vfill \eject

Fig. 9. The standard Ponzano--Regge model (Hasslacher--Perry
interpretation) as an example of the utilization of the pregeometry
construct methodology (PCM). \vfill \eject

Fig. 10. A flat (zero volume) tetrahedron with a dihedral angle along
the edge $\bar{AC}$ equal to $\pi$ radians. \vfill \eject

Fig. 11. The graphical representation of the inequalities in expressions
(\ref{eq-ineq11}) and (\ref{eq-ineq12}). \vfill \eject

Fig. 12. A flat (zero--volume) tetrahedron with a dihedral angle along
the edge $\bar{AC}$ equal to zero radians. \vfill \eject

Fig. 13. The graphical representation of the inequalities in expressions
(\ref{eq-ineq21}) and (\ref{eq-ineq22}). \vfill \eject

Fig. 14. The diagrammatic representation of Step 1 in our construction of
$4$--geometry from the Ponzano--Regge spin networks. Each tetrahedron of
the original lattice is subdivided into four tetrahedra via an identity
for $6j$--symbols. \vfill \eject

Fig. 15. The diagrammatic representation of Step 2 in our construction of
$4$--geometry from the Ponzano--Regge spin networks. Each pair of
tetrahedra sharing a triangle of the original lattice (before Step 1 was
implemented) is subdivided into three tetrahedra via the
Elliot-Biedenharn identity. \vfill \eject

Fig. 16. The diagrammatic representation of Step 3 in our construction of
$4$--geometry from the Ponzano--Regge spin networks. Subdivision of the
$n$ tetrahedra sharing an edge of the original tetrahedral lattice, via
the triangulation of the polygon dual to, and pierced by, the edge,
produces $2n-3$ tetrahedra in their place. This is just a repeated
application of the Elliot--Biedenharn identity using new edges which
pierce triangles which are not part of the original lattice. \vfill
\eject

Fig. 17. The null--strut lattice. The $4$--dimensional blocks in the thin
null--strut sandwich ($TET$,$P-TET^{\ast}$ reciprocal) for the closed
quantity production lattice. They are, from left to right, {\it the
wigwam $4$--simplex} built of a spacelike tetrahedron in $TET$ (thick
lines), a vertex in the $TET^{\ast}$ reciprocal lattice and four null
struts (thin lines); {\it the wedge simplex filler} built of a spacelike
triangle in $TET$, a spacelike edge in the $TET^{\ast}$ reciprocal
lattice, and six null struts; {\it the backward--reaching filler with
polygonal base} built of a spacelike edge of $TET$, a rigidified
(triangulated) spacelike polygon in the $TET^{\ast}$ reciprocal lattice,
and $i$ null struts, where $i$ is the number of vertices of the polygon
(hexagon shown); and {\it the fluted cone} built of a vertex in $TET$, a
spacelike polyhedron with triangulated surface polygons in the
$TET^{\ast}$ reciprocal lattice, and $j$ null struts, where $j$ is the
number of vertices of the polyhedron (truncated octahedron shown). \vfill
\eject

Fig. 18. The use of spin properties to build two, three, and
four--dimensional lattice structures. \vfill \eject

Fig. 19. The Ponzano--Regge model reinterpreted as an example of the
utilization of the pregeometry construct methodology (PCM) for the
construction of a $4$--dimensional pregeometric model. \vfill \eject

Fig. 20. The evolution of a hypersurface by different choices of
foliation (dotted vs dashed slices).

\end{document}